\documentclass{INTERSPEECH2023}

\usepackage{amsmath,graphicx}
\usepackage{url}
\usepackage{hyperref}
\usepackage{multirow}
\usepackage{stfloats}

\fnbelowfloat 

\interspeechcameraready

\title{Text-only domain adaptation for end-to-end ASR using integrated text-to-mel-spectrogram generator}
\name{
    \begin{tabular}{c} Vladimir Bataev$^{1,2}$, Roman Korostik$^{1,3}$, Evgeny Shabalin$^{1,4}$, Vitaly Lavrukhin$^1$, Boris Ginsburg$^1$
    \end{tabular}
}
\address{
  $^1$NVIDIA \\ 
  $^2$University of London, UK; \ \ \
  $^3$ITMO University, Russia; \ \ \ 
  $^4$Higher School of Economics, Russia
}
\email{\{vbataev, rkorostik, eshabalin, vlavrukhin, bginsburg\}@nvidia.com}

\begin{document}


\maketitle
 
\begin{abstract}
We propose an end-to-end Automatic Speech Recognition (ASR) system that can be trained on transcribed speech data, text-only data, or a mixture of both. The proposed model uses an integrated auxiliary block for text-based training. This block combines a non-autoregressive multi-speaker text-to-mel-spectrogram generator with a GAN-based enhancer to improve the spectrogram quality. The proposed system can generate a mel-spectrogram dynamically during training. It can be used to adapt the ASR model to a new domain by using text-only data from this domain. We demonstrate that the proposed training method significantly improves ASR accuracy compared to the system trained on transcribed speech only. It also surpasses cascade TTS systems with the vocoder in the adaptation quality and training speed.
\end{abstract}

\section{Introduction}

\label{sec:intro}

Modern end-to-end automatic speech recognition (ASR) systems are superior to traditional HMM-DNN systems in both academic benchmarks and commercial applications~\cite{li2022recent}.
But customizing such models, especially adapting to a new domain, is still challenging. The standard approach of using audio-text pairs from a new domain has a high cost of collecting and transcribing speech. In many scenarios, text-only adaptation is preferable due to the wide availability of text data. In traditional HMM-DNN systems, a WFST recognition graph is built using a statistical language model (LM), which can be easily constructed from the new text-only data and can significantly improve the performance on a new domain. 
End-to-end ASR systems can also benefit from an external language model~\cite{laptev22ctc}, but this approach requires significantly more computational resources than greedy decoding.

To use text data for training or finetuning, audio can be synthesized from the text to use in a conventional training or finetuning pipeline~\cite{li2018trainingnv, laptev2020you, li2020developingrnnt, zheng2020usingsynaudio, Robinson2022WhenIT}
or the unpaired text input itself can be integrated into an end-to-end ASR neural system~\cite{Sainath2020AnAJ, thomas2022integrating, chen2022maestro, sato2022text, mittal2023insitu}.
Furthermore, some works propose modifying end-to-end model architectures to make the decoder behave like an actual language model for encoder-decoder with attention~\cite{Pham2019IndependentLM} and RNN-Transducer~\cite{meng2022mhat} models, and by such behavior enable text-only adaptation for the decoder part. 
These approaches require changing the training pipeline significantly, in most cases making it incompatible with existing models. Using a pretrained text-to-speech (TTS) system to synthesize audio for audio-text pairs is simpler and does not require modification of existing models.

The main disadvantages of TTS-based ASR customization are a requirement for large storage for generated data, the high computational cost of generating speech, and a mismatch between natural and synthetic audio. By generating speech on the fly using a multi-speaker TTS system, it is possible to produce a practically infinite amount of data without limitations of required space. Each text can be generated with different speakers, and the number of speakers can also be unlimited by sampling from speaker embedding space~\cite{Stanton2021SpeakerG}. Conversely, dynamic audio generation using the complete TTS system will significantly slow down the training process and impact used memory.

Modern TTS pipelines are usually composed of a mel spectrogram generator followed by a vocoder to transform the spectrogram into a speech signal, and the vocoder is usually the most computationally intensive part of the TTS pipeline. Since ASR models can also use mel spectrogram features as input, the vocoder part can be omitted. On the other side, training on spectrograms computed from vocoder output leads to higher ASR performance~\cite{laptev2020you,ueno2022phoneinf}.

Some solutions are proposed to solve the mismatch problem, e.g., training ASR and TTS models with shared components~\cite{Karita2019SemisupervisedES} or with consistency loss~\cite{hori2019cycle, wang2020imrpoving, baskar2021eat, Chen2022Tts4pretrain2A}, using discrete representations instead of mel spectrogram~\cite{ueno2021dataaugtts} or training an additional input block for the combination of ASR and TTS system to improve performance with a frozen ASR backbone~\cite{kurata2021improving}. Most of these methods make ASR and TTS models dependent on each other, and we are trying to avoid this. The rejection sampling algorithm can be used to improve distribution mismatch~\cite{hu2021syntplusplus}, but this exacerbates the problem of computing resources. Also, recently an approach of using an additional block for text-to-mel-spectrogram generator was presented in~\cite{ueno2022phoneinf} with 5-10\% relative WER improvements.

In our work, we are building a lightweight modular text-only adaptation system based on a text-to-mel-spectrogram generator\footnote{\begin{scriptsize}\vspace{-1pt}We limit the investigated task to adaptation to a new text domain. We are \vspace{-1pt}building a modular system, therefore training ASR and TTS models together is \vspace{-1pt}undesirable. We also assume that the model will be applied in the target domain \vspace{-1pt}and do not mitigate performance drop on the source domain.\end{scriptsize}}.  
We augment an end-to-end ASR model with an additional module for the on-the-fly generation of spectrograms from the text during training.
We use FastPitch~\cite{lancucki2021fastpitch} modified to produce synthetic mel spectrograms with the same STFT parameters that are used in the ASR front-end. This approach does not significantly affect the training speed and is fully compatible with existing pretrained models. 
We address the problem of the mismatch between generated and real spectrograms by applying a small GAN-based enhancer directly to the generated spectrograms.
Our main contributions are:
\begin{itemize}
  \item We extend the text-to-spectrogram (TTS) module with an additional block based on StyleGAN 2 to mitigate the mismatch between real and synthetic spectrograms. 
  \item We show that fine-tuning the ASR model on combined speech-text and text-only datasets using the proposed system leads to a significant accuracy improvement. We also demonstrate that one can improve the acoustic end-to-end ASR model by utilizing textual data, usually used for training a language model in hybrid HMM-DNN models.
\end{itemize}

\ifinterspeechfinal
     Models and code are released in the NeMo~\cite{kuchaiev2019nemo} framework.
\else
     Models and code will be available publicly.
\fi

\section{ASR Model with integrated text-to-spectrogram front-end}
\begin{figure}[t]
\centering
\includegraphics[scale=0.8]{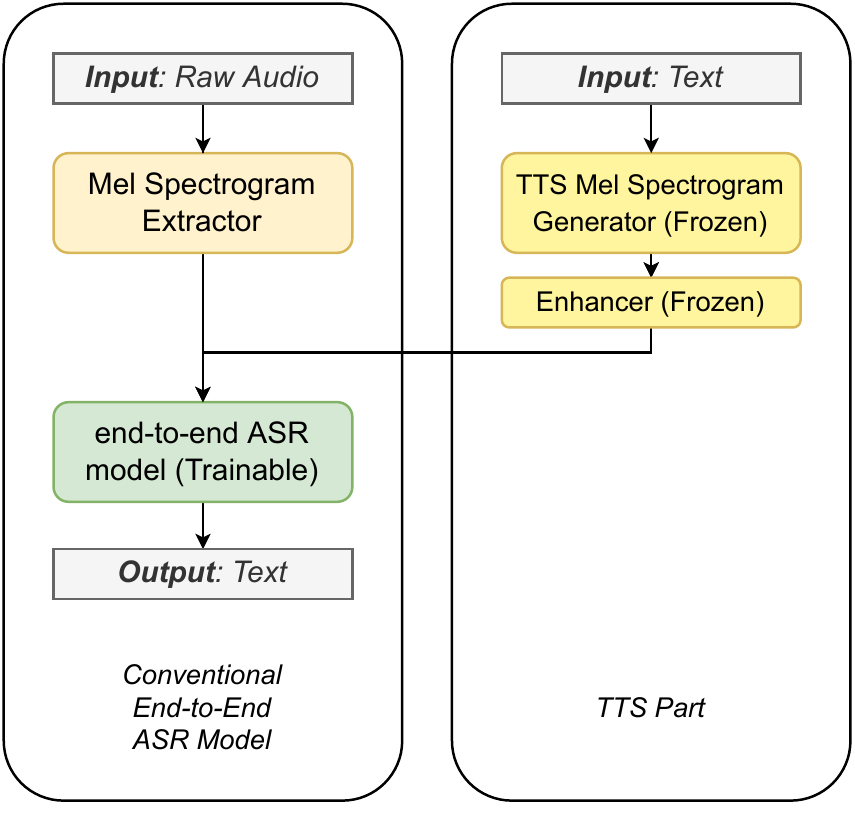}
\caption{ASR system with text-to-mel-spectrogram frontend}
\label{fig:asr-tts-model}
\vspace{-5pt}
\end{figure}

To enable text-only adaptation, we add a text-to-spectrogram front-end to the ASR model as shown in Fig.~\ref{fig:asr-tts-model}. 
The model can take text or audio as input during training.
If the input is audio, the mel spectrogram is extracted and passed directly to the ASR network.
If the input is text, the mel spectrogram is produced on the fly using the pretrained frozen mel spectrogram generator and then fed into the ASR model, as in the previous case.
At inference time, only the standard speech front-end is used. 

\subsection{Text-to-spectrogram frontend}
\label{sec:tts}
We use multi-speaker FastPitch~\cite{lancucki2021fastpitch} as our base text-to-spectrogram model. 
Neural TTS models produce blurry spectrograms, so we add an Enhancer block to add fine details to synthesized spectrograms. Our Enhancer is based on the StyleGAN 2~\cite{karras2020analyzing} architecture\footnote{As a starting point, we used StyleGAN 2 implementation from \begin{scriptsize}\url{https://github.com/lucidrains/stylegan2-pytorch}\end{scriptsize}}.

We modify the original StyleGAN 2 to operate on 80-band mel spectrograms of arbitrary length $L$, which are treated as grayscale images. The network starts from a $5*L/16$ fixed random image and outputs an $80*L$ detailed spectrogram. We add an appropriately downscaled and broadcasted spectrogram to the input and output of each generator block. This way, the generative process for the residual becomes spatially conditioned on the input spectrogram, and the network learns only to generate the details. Within the discriminator, we average across the time axis before projecting to logits.

\begin{figure}[t]
\centering
\includegraphics[scale=0.8]{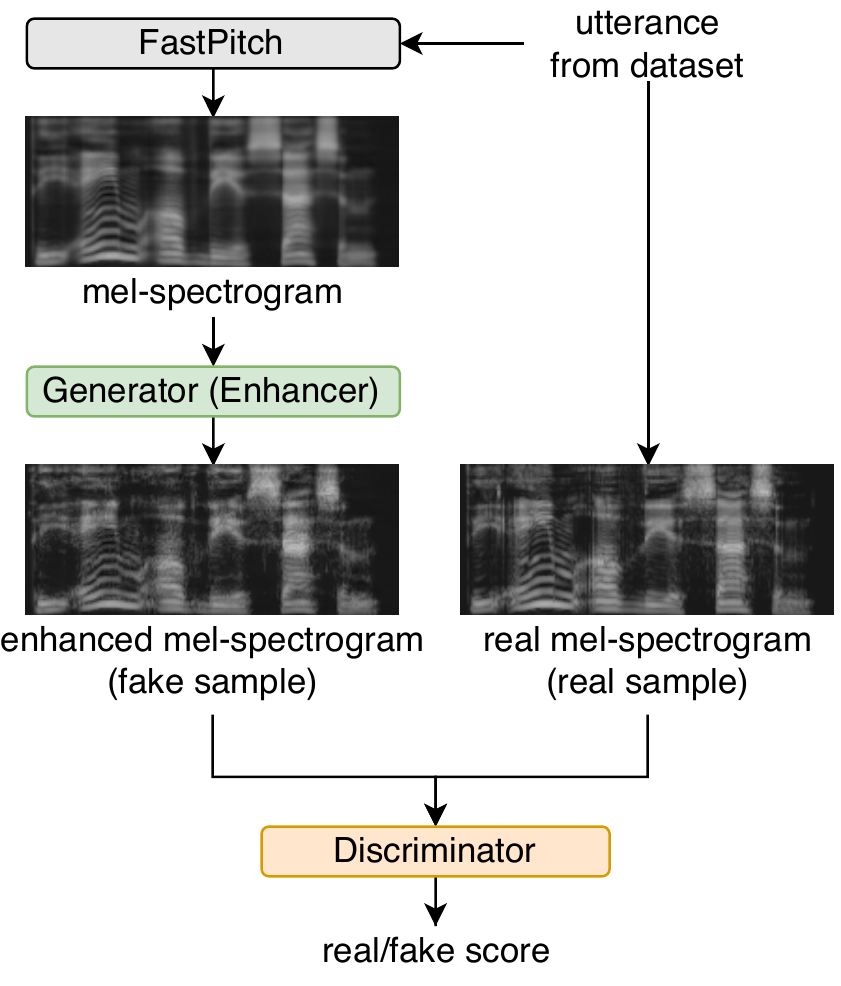}
\caption{Spectrogram Enhancer adversarial training setup}
\label{fig:sg2-training}
\vspace{-5pt}
\end{figure}

The Enhancer is trained adversarially. We generate blurry spectrograms by passing TTS training data through the corresponding FastPitch model using ground-truth $F_0$ and speaker IDs. Blurry spectrograms passed through the generator are considered ``fake'' and real spectrograms are considered ``real''
(Fig.~\ref{fig:sg2-training}).
The training procedure is the usual alternation between gradient steps for the discriminator and the generator. Both are trained using hinge loss~\cite{lim2017geometric}.
The gradient penalty loss~\cite{mescheder2018training, karras2020analyzing} is used every 4 steps for the discriminator.

Ablation runs showed that the enhancer is prone to generating artifacts. To fix this, we add a consistency loss during generator training: L1 distance between real and fake spectrograms, both downsampled 4x along the frequency axis. 

\subsection{ASR backbone}
\label{sec:asr-backbone}
In the proposed pipeline, any end-to-end model can be used as an ASR backbone, which takes in a spectrogram and produces the corresponding transcription. For experiments we use Conformer-L~\cite{gulati2020conformer} (120M), and Conformer-M (32M parameters) Transducers \footnote{\begin{scriptsize}\url{https://github.com/NVIDIA/NeMo/blob/main/examples/asr/conf/conformer/conformer_transducer_bpe.yaml}\end{scriptsize}} implemented with NeMo~\cite{kuchaiev2019nemo}. 
The encoder is a convolution-augmented transformer network; the decoder is a single-layer LSTM network with 640 hidden units. The encoder contains either BatchNorm (BN)~\cite{ioffe2015batch} layers following the original architecture, or LayerNorm (LN)~\cite{ba2016layer}.
Since BN layers in inference mode use statistics accumulated at training time, due to the mismatch between synthetic and real audio additional issues are introduced when using synthetic audio for finetuning. 
To avoid this, we find it useful either to use models with LayerNorm, which normalizes the input samples independently and doesn't have a mismatch between inference and training, or to fuse the BatchNorm layer (Fused BN) into a trainable projection. 
 This is equivalent to removing BN (it can be further fused with a convolutional layer), but it allows us to use pretrained BN-based models for adaptation. Separating statistics in BN for real and synthetic data was proposed in~\cite{hu2021syntplusplus}, but this approach can be applied only when the real data is present. We believe that fusing BN is a more universal alternative for finetuning since we did not observe any meaningful quality degradation between pure and fused BN even for real audio. 

\begin{table}[t]
\caption{\label{tab:benchmark}\textbf{Training overhead} for Conformer-M on text vs real audio. Relative average time per training batch, measured on synthetic LibriSpeech data, NVIDIA V100 GPU.}
\vspace{-13pt}
\begin{center}
\begin{tabular}{l c c }
 \textbf{Model} & \textbf{Input} & \textbf{Relative Training Time} \\ 
 \toprule
 \midrule
 Conformer-M & Audio & 1x \\
\ \ + FastPitch & Text & 1.03x \\
\ \ + FastPitch + Enhancer & Text & 1.10x \\
\ \ + FastPitch + Vocoder & Text & 1.35x \\
 \bottomrule
\end{tabular}
\end{center}
\vspace{-10pt}
\end{table}

\section{Experiments}
\label{sec:exp}

\subsection{Experimental setup}
Conformer was trained and finetuned with the AdamW~\cite{loshchilov2017decoupledadamw} optimizer with a weight decay of $10^{-3}$. In training from scratch, we use the Noam annealing learning rate (LR) scheduler~\cite{vaswani2017attention} and global batch size of $2048$ and train the model for $600$ epochs (1920 GPU hours). For finetuning, we use the Cosine Annealing~\cite{loshchilov2016sgdrcos} LR scheduler with a warmup of 20\% of training steps, and maximum LR of $10^{-4}$. We finetune Conformer-Medium for 10K (290 GPU hours without enhancer) and Large for 15K steps (420 GPU hours) on SLURP~\cite{bastianelli2020slurp} data with global batch size of 512. For finetuning with WSJ~\cite{paul1992wsj} data we use batch size of 256, and finetune for 20K steps on usual training data (160 GPU hours), and for 40K when mixing texts and audio-text pairs. In all scenarios SpecAugment~\cite{park2019specaugment} is applied for both real and synthetic data. All experiments are done on the cluster of NVIDIA DGX Stations. Conformer models are trained on 4 nodes, each containing 16 V100 GPUs.

For setup with Enhancer, we train FastPitch using NeMo toolkit~\cite{kuchaiev2019nemo} with configuration \texttt{fastpitch\_align\_v1.05} (50.7 M parameters).
The Enhancer has a latent dimension of $192$, depth of the style network is $4$, ``network capacity'' is 16, and the upper bound on the number of feature maps in convolutional layers is $192$. The generator has 3.5M parameters, and the discriminator has 4.5M parameters. We trained model on 8 V100 GPUs for 20 epochs using Adam~\cite{DBLP:journals/corr/KingmaB14} optimizer with $\beta_1 = 0.5$, $\beta_2 = 0.9$ and LR of $2 \cdot 10^{-4}$. Batch size is $128$, and the consistency loss has a weight of $0.1$. We use greedy decoding for WER evaluation. All experiments done on LibriTTS train-clean-100 (53 hours) and train-960 (585 hours) subsets. After filtering out utterances longer than 15s, we are left with 45 hours and 468 hours respectively. Spectrogram parameters follow ASR: 25ms window, and 10ms hop, with mel bands ending at 8 kHz.

For the text-to-waveform TTS setup we use FastPitch with UnivNet~\cite{jang2021univnet} vocoder trained on LibriTTS clean-100 subset. Intermediate acoustic features are 80-band mel-spectrograms, calculated using 1024-sample windows with 256-sample hop calculated from a 22.05 kHz signal. Mel bands end at 8 kHz. Effectively this means $\sim$46ms window size with $\sim$12ms hop.

The Enhancer-based TTS frontend is much faster than the Conformer models. During training, we observe an overhead of $\leq 10 \%$ for Conformer-M (Table~\ref{tab:benchmark}). The benchmarking was done on V100 GPU with batch size of 16. 

\subsection{Training ASR model using synthetic only data}

\begin{table}[t]
\caption{\label{tab:train-ls-synthetic}\textbf{Text-only training on LibriSpeech texts.} Model: Conformer-M with LN. Greedy WER[\%].}
\vspace{-13pt}
\begin{center}
\begin{tabular}{ l c c c c c }
\multirow{2}{10em}{\textbf{Training setup}} & \multicolumn{2}{c}{\textbf{dev}} & \multicolumn{2}{c}{\textbf{test}} \\ 
 & \textbf{clean} & \textbf{other} & \textbf{clean} & \textbf{other} \\
 \toprule
 \midrule
 TTS 45h & 27.0 & 59.2 & 27.6 & 61.8 \\
 \ \ + vocoder & 16.3 & 44.5 & 17.3 & 46.2 \\
 \ \ + enhancer & 11.6 & 38.9 & 12.6 & 40.8 \\
 \midrule
 TTS 468h & 16.5 & 43.2 & 17.1 & 45.5 \\
 \ \ + enhancer & 8.9 & 22.3 & \textbf{9.3} & \textbf{22.1} \\
 \midrule
\multicolumn{1}{ l }{Oracle (real audio-text)} & 2.7 & 6.6 & 2.9 & 6.6 \\ 
\bottomrule
\end{tabular}
\end{center}
\vspace{-10pt}
\end{table}

We start our experiments with a synthetic but representative setup, to demonstrate the problem of mismatch between real and synthetic data for training ASR models. In such a setup, a model trained only on real audio typically fails to generalize, requiring a mixture of real and synthetic audio to achieve good performance~\cite{li2018trainingnv}.
We pretrain the FastPitch using subsets of LibriTTS~\cite{zen2019libritts}, which consists of the same speakers and mostly the same data as LibriSpeech.
We then only use text from the LibriSpeech~\cite{panayotov2015librispeech} audio-text paired dataset to generate spectrograms with random speakers, to train a Conformer-M model with LayerNorm.
We can see that using more data for TTS model leads to a significantly lower WER on both "clean" and "other" parts of the dataset (see Table~\ref{tab:train-ls-synthetic}), but the difference compared to real audio (6.3\% vs 45.5\% WER on test-other) is still considerable. Using the enhancer model to generate more realistic spectrogram allows to achieve significantly better results (22.1\% WER on test-other), and also surpasses cascade TTS system with vocoder.

\subsection{Text-only finetuning on SLURP dataset}
\begin{table}[t]
\caption{\label{tab:finetune-ls-slurp}\textbf{Finetuning model with LayerNorm on SLURP text-only data.} Base model: Conformer-M with LN, trained on LibriSpeech. Greedy WER[\%].
}
\vspace{-13pt}
\begin{center}
\begin{tabular}{ l c c c }
\textbf{Model} & \textbf{dev} & \textbf{test} \\ 
\toprule
\midrule
 base: Conformer-M with LN & 48.9 & 49.3 \\
 \midrule
 \ \ + SLURP texts $\rightarrow$ TTS 45h & 37.5 & 38.1 \\
 \ \ \ \ \ \ + vocoder & 36.9 & 37.9 \\
 \ \ \ \ \ \ + enhancer & 36.1 & 36.8 \\
  \midrule
 \ \ + SLURP texts $\rightarrow$ TTS 468h & 33.9 & 34.7 \\
 \ \ \ \ \ \ + enhancer & \textbf{31.3} & \textbf{32.4} \\
 \bottomrule
\end{tabular}
\end{center}
\vspace{-10pt}
\end{table}

\begin{table}[t]
\caption{\label{tab:finetune-ls-slurp-batchnorm}\textbf{Finetuning models with BatchNorm on SLURP text-only data.} Base model: Conformer-M with BN, trained on LibriSpeech. Greedy WER[\%].}
\vspace{-13pt}
\begin{center}
\begin{tabular}{ l c c c }
 \textbf{Model} & \textbf{dev} & \textbf{test} \\ 
 \toprule
 \midrule
 base: Conformer-M with BN & 49.6 & 49.8 \\
 \midrule
 \ \ + SLURP texts & 36.6 & 37.5 \\
 \ \ \ \ \ \ + fused BN & 34.2 & 35.1 \\ 
 \ \ \ \ \ \ + enhancer & 33.1 & 33.9 \\
 \ \ \ \ \ \ + enhancer + fused BN & \textbf{31.9} & \textbf{32.5} \\ 
 \bottomrule
\end{tabular}
\end{center}
\vspace{-10pt}
\end{table}

We use the SLURP~\cite{bastianelli2020slurp} dataset to study text-only adaptation using the proposed approach. SLURP is a spoken language understanding dataset, which contains diverse verbose commands for smart home control. Since its text significantly differs from audio books and conventional dialog data, it is suitable for investigating text-only adaptation.

Base models are trained on LibriSpeech data, and we use text-only data from SLURP training set ($\sim$11K utterances) to adapt the model. We report WER for the original dev and test subsets containing each $\sim$10h of text-audio pairs.
Similar to the previous setup, ablation studies in Table~\ref{tab:finetune-ls-slurp} with finetuning Conformer-M with LN show improvement from using more data for mel spectrogram generator (train-clean-100 vs full LibriTTS), and improvement from the enhancer. 

Finetuning models with BN is more challenging. As described in Section \ref{sec:asr-backbone}, we replace BN with a trainable projection initialized from the original layer parameters. Experiments in Table~\ref{tab:finetune-ls-slurp-batchnorm} show that finetuning Conformer-M with fused BN on text-only data leads to larger WER improvements. The relative impact of the enhancer is larger on models with pure BN, but the system with fused BN has better performance and is comparable with LN-based models.

Table~\ref{tab:finetune-ls-slurp-large-vs-medium} shows that larger ASR model size leads to better relative improvement, and the system trained on LibriSpeech with the baseline WER of 47.1\% is able to achieve 27.7\% WER on SLURP test (41.2\% relative improvement).

\begin{table}[t]
\caption{\label{tab:finetune-ls-slurp-large-vs-medium} \textbf{Finetuning medium and large models on SLURP text-only data}. Base models: Conformer-M and Conformer-L with fused BN, trained on LibriSpeech. Greedy WER[\%].}
\vspace{-13pt}
\begin{center}
\begin{tabular}{ l c c c }
 \textbf{Model} & \textbf{dev} & \textbf{test} \\ 
\toprule
base: Conformer-M & 49.6 & 49.8 \\
 \ \ + SLURP texts + enhancer & 31.9 & 32.5 \\
 \midrule
 base: Conformer-L & 46.7 & 47.1 \\
 \ \ + SLURP texts + enhancer & \textbf{27.0} & \textbf{27.7} \\ 
 \bottomrule
\end{tabular}
\end{center}
\vspace{-15pt}
\end{table}

\begin{table}[b]
\vspace{-5pt}
\caption{\label{tab:finetune-asrset3-slurp}\textbf{Finetuning model pretrained on large ASR corpus on SLURP text-only data.} Base: Conformer-L with BN, trained on a large (24K hours) audio-text dataset. Greedy WER[\%]}
\vspace{-13pt}
\begin{center}
\begin{tabular}{ l c c c }
 \textbf{Model} & \textbf{dev} & \textbf{test} \\ 
 \toprule
 \midrule
 base: Conformer-L & 23.2 & 23.5 \\
  \midrule
 \ \ + SLURP texts + enhancer & 16.1 & 16.7 \\ 
 \ \ \ \ \ \ + fused BN & \textbf{14.7} & \textbf{15.0} \\ 
 \bottomrule
\end{tabular}
\end{center}
\vspace{-12pt}
\end{table}

Text-only adaptation helps even when the ASR model has been trained on a large amount of speech data. For this scenario we finetuned Conformer-L pretrained on a large corpus of 24K hours of read and conversational speech data (not including SLURP) starting from a publicly available checkpoint\footnote{\begin{scriptsize}\url{https://catalog.ngc.nvidia.com/orgs/nvidia/teams/nemo/models/stt_en_conformer_transducer_large}\end{scriptsize}}. 
With text-only finetuning the system is able to achieve 15.0\% WER (36.2\% relative improvement), see Table~\ref{tab:finetune-asrset3-slurp}.

\subsection{Finetuning on WSJ dataset}
To compare the contributions of using original audio-text pairs and synthetic data whilst using a large text corpus for improving the model, we use the WSJ~\cite{paul1992wsj} corpus. The audio conditions are similar to LibriSpeech, however the domain of text is different. 
To compare the text-only adaptation with text-audio pairs we use text-only training data. The results are in Table~ \ref{tab:finetune-wsj}.
Our system achieves 47\% relative improvement on eval-92 using only texts, and surpasses text-only adaptation of MHAT~\cite{meng2022mhat}, which achieves 24\% WERR (from 9.5\% to 7.2\%) in the same scenario.
The quality of text-only adaptation is still below than the quality of training on real data, but the difference is small.
Also, we observe that the proposed method can not only replace adaptation on real data with using text-only samples, but it can gain benefits by combining them both. WSJ contains also a large LM corpus, used in traditional HMM-DNN systems to build a speech recognition graph based on a statistical LM. When using synthetic mel spectrogram from this text, we observe significant improvement compared to using speech-only data, achieving 1.5\% WER on eval-92 (Table~\ref{tab:finetune-wsj-lm-text}). Each training epoch combines speech data and sampled texts from the LM corpus with the shown ratio.

\begin{table}[t]
\caption{\label{tab:finetune-wsj} \textbf{Finetuning on WSJ: audio-text pairs vs text-only data.} Base model: Conformer-L with fused BN trained on LibriSpeech. Greedy WER[\%]}
\vspace{-13pt}
\begin{center}
\begin{tabular}{ l c c }
 \textbf{Model} & \textbf{dev-93} & \textbf{eval-92} \\ 
 \toprule
 \midrule
base: Conformer-L & 9.3 & 7.2 \\
 \midrule
 texts & 5.5 & 3.8\\
 audio-text pairs & \textbf{4.2} & \textbf{3.1} \\
 \bottomrule
\end{tabular}
\end{center}
\vspace{-10pt}
\end{table}

\begin{table}[t]
\caption{\label{tab:finetune-wsj-lm-text}\textbf{Finetuning using WSJ audio-text pairs and text from WSJ Language Model corpus.} Base model: Conformer-L with fused BN trained on LibriSpeech. Greedy WER[\%]. }
\vspace{-13pt}
\begin{center}
\begin{tabular}{ c c c c }
 \textbf{audio:text ratio} & \textbf{dev-93} & \textbf{eval-92} \\
\toprule
\midrule
 base: Conformer-L & 9.3 & 7.2 \\
 \midrule
 1:1 & 2.7 & 1.7 \\
 1:2 & \textbf{2.5} & \textbf{1.5} \\
 \bottomrule
\end{tabular}
\end{center}
\vspace{-15pt}
\end{table}

\section{Conclusions}
\label{sec:conclusions}
In this paper, we presented an end-to-end ASR model which can be trained both on text-audio pairs and text-only data. We enable text-only training using an integrated neural text-to-spectrogram module, composed of a modified non-autoregressive text-to-mel-spectrogram generator with a lightweight GAN-based spectrogram enhancer. 
Such a model does not require external storage for generated audio and also outperforms the TTS system with a vocoder in both speed and quality. The amount of generated synthetic data can be theoretically infinite when using a multi-speaker text-to-mel-spectrogram model.
The Enhancer helps to mitigate the mismatch between real and synthetic mel spectrograms. This leads to significant improvement of ASR accuracy in text-only adaptation.

By using text-only LibriSpeech data to train ASR model from scratch, we show the effectiveness of the proposed approach. We also show up to the 41.2\% relative improvement on SLURP data in the text-only scenario. On the WSJ dataset, we demonstrate the effectiveness of our proposed system by combining a large text corpus with text-audio pairs and achieving 1.5\% WER on the test set.

We used Conformer as ASR backbone and FastPitch as text-to-mel-spectrogram generator. Since the proposed architecture is modular, the approach should generalize to other ASR and TTS models.

\ifinterspeechfinal
\section{Acknowledgments}
\label{sec:acknowledgments}
We thank our colleagues Aleksandr Laptev, Sean Narenthiran, Jocelyn Huang, and Elena Rastorgueva for the help with the code and paper review.
\fi


\bibliographystyle{IEEEtran}
\bibliography{refs}

\end{document}